\documentclass[referee]{raa}
\usepackage{graphicx,times}             
\usepackage{natbib}
\usepackage{amssymb,amsmath}
\bibpunct{(}{)}{;}{a}{}{,}

\usepackage[a4paper=true,dvipdfm=true,pagebackref=true]{hyperref}
\hypersetup{colorlinks = true, linkcolor = green, anchorcolor = red, citecolor = blue, filecolor = red, pagecolor = red, urlcolor = red}

\newcommand{\teff} {\mbox{\rm $T_{eff}$}}

\newcommand{\feh} {\mbox{\rm [Fe/H]}}

\newcommand{\afe} {\mbox{\rm [$\alpha$/Fe]}}
\newcommand{\mgfe} {\mbox{\rm [Mg/Fe]}}
\newcommand{\alfe} {\mbox{\rm [Al/Fe]}}
\newcommand{\sife} {\mbox{\rm [Si/Fe]}}
\newcommand{\cafe} {\mbox{\rm [Ca/Fe]}}
\newcommand{\tife} {\mbox{\rm [Ti/Fe]}}

\newcommand{\cfe} {\mbox{\rm [C/Fe]}}

\newcommand{\ofe} {\mbox{\rm [O/Fe]}}

\newcommand{\cn} {\mbox{\rm [C/N]}}

\newcommand{\Vphi} {\mbox{\rm $V_{\rm{\phi}}$}}

\newcommand{\kmprs} {\mbox{\rm \,$km/s$\,}}

\begin{document}
\title{High-$\alpha$-Metal-Rich Stars in the LAMOST-MRS Survey and Its Connection with the Galactic Bulge}


\volnopage{Vol.0 (20xx) No.0, 000--000}      
\setcounter{page}{1}          

\author{Haopeng Zhang
      \inst{1,2}
   \and Yuqin Chen
      \inst{1,2}
   \and Gang Zhao
      \inst{1,2}
   \and Jingkun Zhao
      \inst{1,2}
   \and Xilong Liang
      \inst{1}
   \and Haining Li  
      \inst{1}
   \and Yaqian Wu
      \inst{1}
   \and Ali Luo  
      \inst{1}
   \and Rui Wang
      \inst{1}
   }


   \institute{CAS Key Laboratory of Optical Astronomy, National Astronomical Observatories, Chinese Academy
of Sciences, Beijing 100101, China; {\it cyq@bao.ac.cn}\\
        \and
             School of Astronomy and Space Science, University of Chinese Academy of Sciences, Beijing 100049, China\\
\vs\no
   {\small Received~~20xx month day; accepted~~20xx~~month day}}

\abstract{
We report the detection of a large sample of high-$\alpha$-metal-rich stars on the low giant branch with $2.6<logg<3.3$ dex in the LAMOST-MRS survey. This special group corresponds to an intermediate-age population of $5-9$ Gyr based on the $\feh$-$\cn$ diagram and age-$\cn$ calibration.  A comparison group is selected to have solar $\alpha$ ratio at super metallicity, which is young and has a narrow age range around 3 Gyr.  Both groups have thin-disk like kinematics but the former shows slightly large velocity dispersions. The special group shows a larger extension in vertical distance toward 1.2 kpc, a second peak at smaller Galactic radius and a larger fraction of super metal rich stars with $\feh>0.2$ than the comparison group. These properties strongly indicate its connection with the outer bar/bulge region at $R=3-5$ kpc. A tentative interpretation of this special group is that its stars were formed in the X-shaped bar/bulge region, close to its corotation radius, where radial migration is the most intense, and brings them to present locations at 9 kpc and beyond.  Low eccentricities and slightly outward radial excursions of its stars are consistent with this scenario. Its kinematics (cold) and chemistry ($\afe\sim 0.1$) further support the formation of the instability-driven X-shaped bar/bulge from the thin disk.
}

\authorrunning{H.P. Zhang et al. }            
   \titlerunning{H$\alpha$mr stars in the LAMOST survey}  

   \maketitle

\section{Introduction}
Stars with super metallicity of $\feh>0$ in the solar neighborhood
and beyond are interesting targets to investigate the role of radial migration 
in the Galactic disk. 
Based on large 
spectroscopic surveys, several works \citep{Kordopatis15,Anders17,Chen19} 
detected hundreds of super metallicity stars in the solar neighborhood.
It has been suggested that these stars originate from the inner Galaxy and 
have moved to present locations by radial migration first
introduced by \cite{Sellwood02}. The Galactic bulge has its main component
with super metallicity \citep{Ness13} and thus becomes the natural
origin of these stars. This connection is also suggested by theoretical 
chemo-dynamical models of \cite{Minchev13} and other simulation works. 
In this respect, the observational properties of stars with super 
metallicity will provide important information on
the Galactic bulge. 

The formation of the Galactic bulge is, however, complicated and has raised hot
debates in recent years. As stated in the review paper by \cite{Barbuy18},
the bulge is previously thought to be a collapsed and rapidly-formed 
spheroidal old component due to early mergers in the Galaxy \citep{Zoccali08}, 
while it has been recently recognized to be a bar formed from dynamical
instabilities in the disk \citep{DiMatteo16}.
The color-magnitude diagrams (CMD) in the bulge region \citep{Zoccali08,Clarkson08} 
would favor an old population of 10 Gyr and the narrow sequence at the 
turn off region of the CMD allows little room for the existence of young stars 
with ages of less than 8 Gyr. By adopting an evolving age-metallicity relation
for stars in the bulge, \cite{Haywood16} reanalyzed
the bulge CMD of \cite{Clarkson08} by isochrones with a spread of ages,
and found that all stars with $\feh>0$ are younger than 8 Gyr, which
would make up  50\% of all the stars in the bulge.
With the release of the infrared spectroscopic data of the APOGEE survey,
more works \citep{Sit20,Bovy19} favor a secular evolution of the bulge 
with a mixing population and thus young and old stars co-exist in the 
inner region of
the Galaxy. In particular, \cite{Bovy19} presented unprecedented detailed
maps of the kinematics, elemental abundances, and ages from the bulge region
to the outer disk. They found a long bar
formed at 8 Gyr ago with a size of 5 kpc, and in the bulge region of $R<5$ kpc
different metallicity and kinematics are shown between stars inside and 
outside the bar, indicating a complicate history of the bulge region. 

In this paper, we investigate metal rich stars in the LAMOST 
mid-resolution spectroscopic survey, which provides 
us with an extraordinary database of radial
velocities, stellar parameters and elemental abundances for
a large number of disk stars. 
In particular, we aim to derive their ages based on age-sensitive abundances,
which is not available in the study of super metallicity stars
from the LAMOST low resolution spectroscopic survey by \cite{Chen19}.
As compared with stellar position and kinematics, age, metallicity
and elemental abundances of metal rich stars bear important
signatures of the bar/bulge population taking into account
the fact that they are not altered by secular evolution, such as 
radial migration and blurring.
Old metal rich stars could be picked out and become the most interesting 
targets to investigate radial migration process of the Galactic disk and 
provide new constraints on the bar/bulge's formation.

\section{Data}
In 2018, the LAMOST Galactic spectroscopic 
survey \citep{Zhao06,Zhao12,Cui12,Deng12}
finished its first five-year regular survey with the low
resolution mode (R$\sim$1800), and started a new five-year
medium resolution ($R\sim 7500$) survey (MRS)
at two wavelength ranges of $4950\AA  < \lambda < 5350\AA$ (Blue band) and
$6300\AA  < \lambda < 6800\AA$ (Red band).
The LAMOST-MRS survey is designed for achieving
several scientific goals, e.g., Galactic archaeology, stellar physics,
star formation, Galactic nebulae, etc \citep{Liu20}. A large
fraction of target stars has multiple observations over three times, which
provide good resulting spectra in order to detect variation in 
radial velocity to exclude binary stars and to select high-quality
spectra in deriving reliable abundances of the sample stars.
The LAMOST-MRS DR7 has released that
2 426 237 spectra have signal-to-noise (S/N) higher than 10 for both
blue and red bands. Recently, \cite{Wang20} provided
radial velocities, stellar parameters
($\teff$, $logg$, $\feh$) and abundances of 12 elements (C, N, O, Mg, Al, Si, Ca, Ti, S, Ni, Cr, Cu)
by using the SPCANet neutral network (hereafter LAMOST-MRS-SPCANet DR7).

Based on this database, we select our sample stars in the following steps.
First, stars on the low giant branch (LGB) 
with $2.6<logg < 3.3$ are picked out because reliable ages can be
estimated from $\cn$ ratios for these stars as shown 
in \cite{Hasselquist19}. Then we limit stars with $\feh > -0.8$
since we focus on the study of metal rich stars in the Galactic disk.
With the above two criteria,
the sample has a temperature range of $4200<\teff < 5200\,K$.
Note that \cite{Wang20} derived abundances for
LAMOST-MRS DR7 sample based on stars in common with
the APOGEE-Payne DR14 database by \cite{Ting19}.
With the release of APOGEE DR16, we compare these abundances
with this more reliable dataset
for stars with $2.6<logg < 3.3$ and derive
the abundance calibrations for Fe, Mg, Al, Si, Ca, Ti, C, N, O elements
based on stars in common by linear fits to the data, which
are shown in red lines of Fig.~1. With these calibrations, we obtain
abundances in the same scale with the APOGEE DR16 dataset.

\begin{figure*}
\includegraphics[scale=1.0]{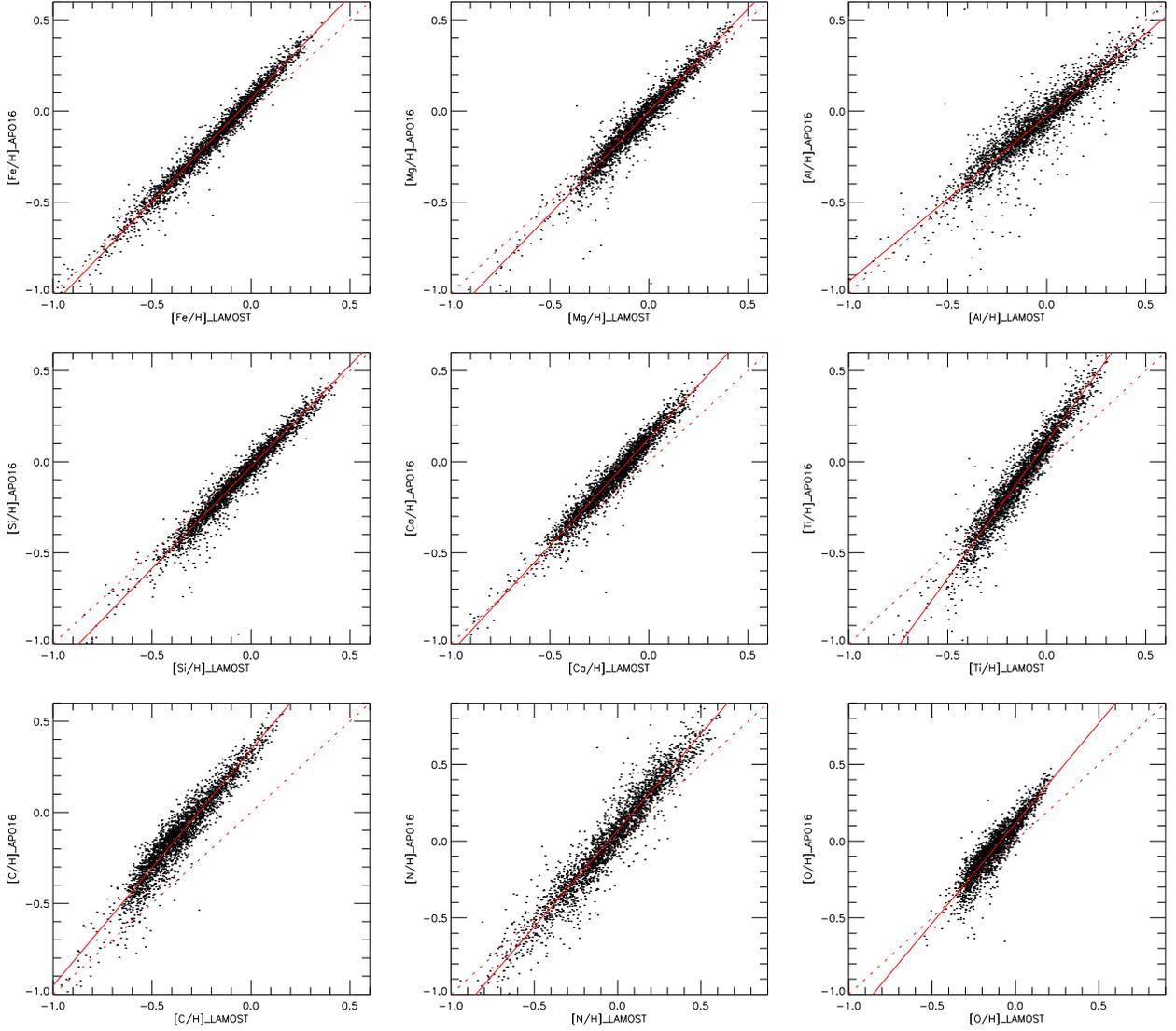}
\caption{The abundance calibrations of $[X/H]$ for Fe, Mg, Al, Si, Ca, Ti, C, N, O elements between the LAMOST MRS DR7 and APOGEE DR16 based on common stars with $2.6 <logg< 3.3$ and $\feh > -1.0$. Red solid lines are linear fits to the data and red dash lines show the one-to-one relations.}
\label{fig1}
\end{figure*}

For the selected sample of stars, we calculate stellar positions,
spatial velocities and orbital parameters
based on radial velocities from the LAMOST-MRS-SPCANet data,
distances and proper motions from \textit{Gaia} DR2 \citep{Brown18}.
The distance of $r_{est}$ from \textit{Gaia} DR2 is adopted and
we limit stars with relative error in parallax to be less than 10\%.
Orbital parameters are calculated
from spatial velocity with the help of Galactic potential
{\it MWPotential2014} in
$Galpy$ by \cite{Bovy15}. Details on the calculations are presented
in \cite{Chen19} for super
metal rich stars in the LAMOST low resolution survey.
The solar radius is
of 8.34 kpc, the solar circular velocity of $240 \kmprs$ \citep{Reid14}
and the solar motion of (11.1, 12.24, 7.25) $\kmprs$ \citep{Schonrich09}.

Fig.~2 shows the
distributions of the sample stars in the $R-Z$ plane.
Here we pick out two specific regions for later comparisons
with blue circles corresponding to stars with $|Z|>2$ kpc
as a representation of the thick disk population and green circles
being stars at the very low Galactic-plane
inner solar circle ($|Z|<0.1$ kpc and $R<8.0$ kpc)
in the LAMOST-MRS-SPCANet DR7 dataset.

\begin{figure}[htpb]
\includegraphics[scale=1.0]{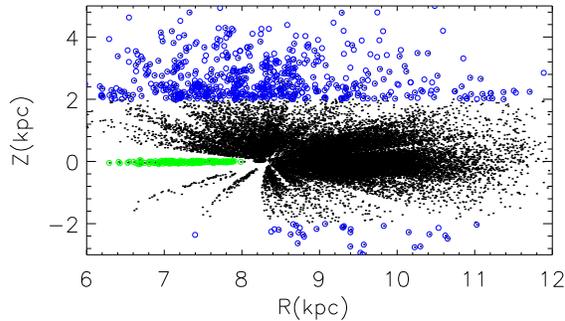}
\caption{The $R-Z$ plane of the selected sample.
For later comparisons, stars at the low-$|Z|$ inner solar circle
with $R<8$ kpc and $|Z|<0.1$ kpc are marked with green circles,
and stars from the thick disk with $|Z|>2$ kpc are marked with blue circles.}
\end{figure}

\section{A special group of stars detected in
the $\feh$ versus $\mgfe$ diagram} 
Fig.~3 shows the $\feh$ versus $\mgfe$
diagram and its contour mapping for our sample stars. 
Two sequences, the thin disk on the low $\mgfe$ branch and
the thick disk on the high $\mgfe$ branch, are well separated by
the red line of $\mgfe=-0.35\feh+0.08$, which is
drawn by passing through the sparse region between the two
branches at $-0.4<\feh<-0.2$.
Interestingly, a special feature with $\feh>0$ 
and $\mgfe\sim 0.1$ is shown in the contour map (lower panel)
and its stars are marked in red dots (upper panel).
We note that this special group is similar as the 
high-$\alpha$-metal-rich (h$\alpha$mr)
stars detected in the solar neighborhood by \cite{Adibekyan12}.
Meanwhile, it could be the same feature as 
the upturn trend of $\sife$ for the most metal rich stars 
in the APOGEE survey by \cite{Haywood13}.
The feature in the present work is clearer than that of the APOGEE data
probably because we limit the sample stars within the LGB stage
of $2.6<logg<3.3$ dex
 so that stars have internally consistent abundances to show 
distinguished features.
Based on 27 135 red clump stars in the APOGEE survey, this group
is clearly picked out as a separated feature by \cite{Ratcliffe20}
(Group 3, their Fig.~11). This supports the reality
of this special group in the LAMOST-MRS survey.

We over-plot stars from the two selected regions (blue and green circles) on
the contour map in the $\feh$ versus $\mgfe$ diagram. It shows
that this special group forms an extension of the thick disk, since
both of them are located above the red line.
However, only one star from the selected thick disk (blue circles)
is found in this region, while quite a few stars from the very low-$|Z|$
inner solar circle (marked by green circles) belong to this special group.
We have 3293 stars in the special group, and we
select a comparison group, which includes 1565 metal rich stars 
with $\feh>0.0$ and $\mgfe<-0.35\feh+0.08$ (below the red line). 
In the following section, we investigate the similarities and differences 
of age, chemical and kinematical properties between the two groups.

\begin{figure}
\includegraphics[scale=1.0]{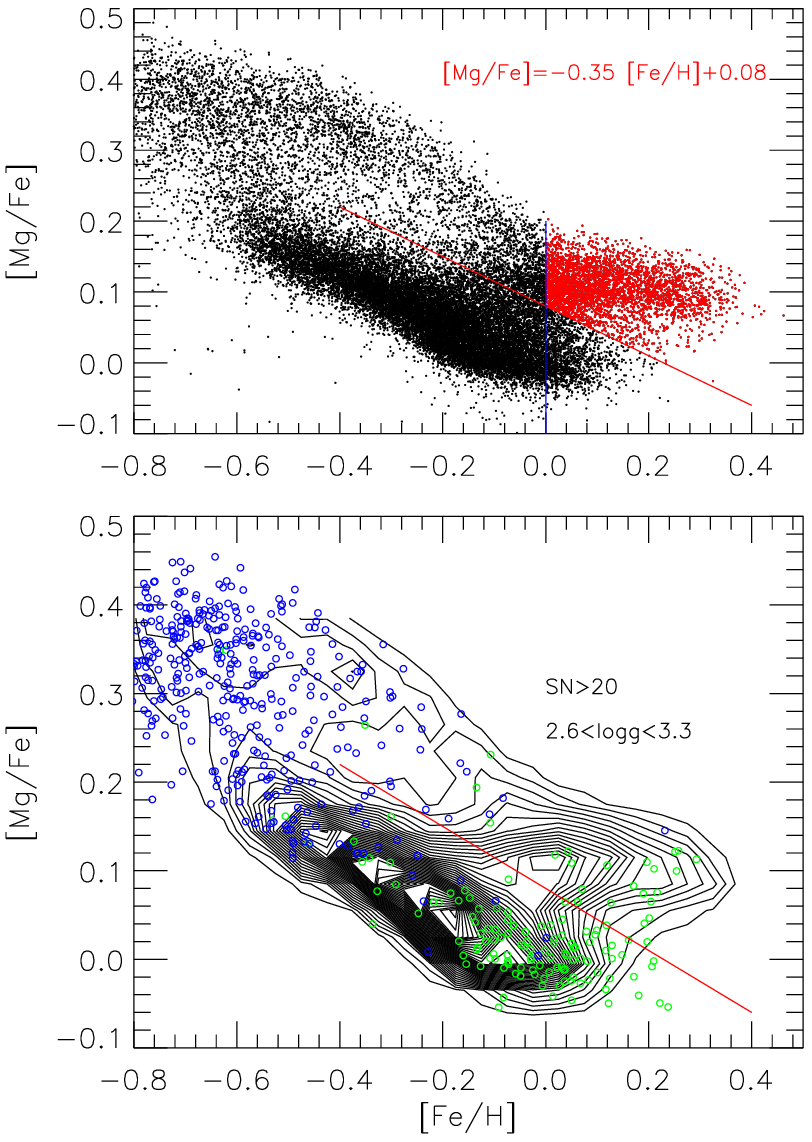}
\caption{The $\feh$ versus $\mgfe$ diagram and its
contour map for our sample stars.
The red line with $\mgfe=-0.35\feh+0.08$ is drawn to separate
the high-$\mgfe$ and low-$\mgfe$ sequences.}
Stars from the low-$|Z|$ inner disk 
at $R<8$ kpc and $|Z|<0.1$ kpc (green circles)
and the thick disk at $2<|Z|<4$ kpc (blue circles) are overplotted.
\end{figure}

\section{The special group versus the comparison group}
\subsection{Ages estimated from $\cn$ ratios}
Based on masses and ages derived by \cite{Martig16} and abundances
from APOGEE DR14, \cite{Hasselquist19}
interpreted the $\cn$ versus $\feh$ distributions
as the age versus $\feh$ trends, and found that
LGB stars with $2.6<logg<3.3$ dex
have the most reliable ages derived from $\cn$ ratios.
Fig.~4 shows the $\feh$ versus $\cn$ diagram 
for our sample stars with red dots corresponding to h$\alpha$mr stars
in the special group.
Two red lines indicate the locations of the old population of $t=9$ Gyr
and the young population of $t=2.5$ Gyr, which are derived from
open clusters in APOGEE DR14 by \cite{Hasselquist19}.
The special group lies below the upper red line
of $t=9$ Gyr but far above the lower red one of $t=2.5$ Gyr,
indicating an intermediate-age population of $5<t<9$ Gyr. 
The comparison group lies around the lower red line with
its main part of $\sim3$ Gyr, slightly above the $t=2.5$ Gyr line.

We estimate ages for our sample stars based on the age calibration
of $log(Age[yrs])=10.54+2.61\cn$ by \cite{Casali19}, which was derived 
from open clusters in the Gaia-ESO and APOGEE surveys.
This calibration is valid for stars with $-0.4<\feh<0.4$,
and our sample stars ($\feh>0$) fits this requirement.
However, the comparison group shows a peak at 1 Gyr by using
this age calibration, which indicates a systematic deviation
in the $\cn$ ratio between \cite{Casali19} and the present work
from the dataset by \cite{Wang20}.
Therefore, we shift the $\cn$ term upward by 0.15 dex
in the age calibration in order to fit
the peak of $2 - 3$ Gyr of the age distribution for the comparison
group. That is, we adopt a revised age calibration of
$log(Age[yrs])=10.54+2.61(\cn+0.15)$.
Note that there is no deviation in the $\cn$ ratio between \cite{Hasselquist19}
and our work because both are based on the APOGEE DR14 dataset.
\cite{Wang20} derived abundances for the
LAMOST-MRS DR7 sample based on stars in common with
the APOGEE-Payne DR14 database by \cite{Ting19}, and thus
it is also based on the APOGEE DR14 dataset.
This consistency seems to persist when we correct abundances 
to the same system of APOGEE DR16 via calibrations in Fig.~1.

Based on the revised age calibration, the special group
shows a wide age range of $5-12$ Gyr with a peak at $7-8$ Gyr.
Since the age calibration of \cite{Casali19} does not take into account
the metallicity dependence, which is clearly seen in open clusters of
\cite{Hasselquist19}, we expect that the old end of the age distribution
is overestimated at the solar metallicity where the highest $\cn$ is
found in the special group.
In view of this, we suggest that this special group
has an upper age limit of 9 Gyr, rather than the $10-12$ Gyr
from the revised age calibration. 

The significant deviation of age by $\sim4-5$ Gyr between the special
and the comparison groups is consistent
with the result of \cite{Sit20} that stars in the bulge are about twice as old
(8 Gyr) as those in the solar neighborhood (4 Gyr) for low-$\alpha$ stars, 
corresponding to a 4-Gyr-old difference.
The age peaking at 7 Gyr of this special group is exactly the same
as $\sim7.5$ Gyr of \cite{Bovy19} (their Fig.~7) for the bar region
based on APOGEE DR16, and is consistent with 
an age of $8\pm3$ Gyr by \cite{Sit20} for bulge stars within
$R<3.5$ kpc in APOGEE DR14.
Thus, the age distribution of the special group indicates
its connection with the bar/bulge, rather than
the local-born young disk in the solar neighborhood.

\begin{figure}
\includegraphics[scale=1.0]{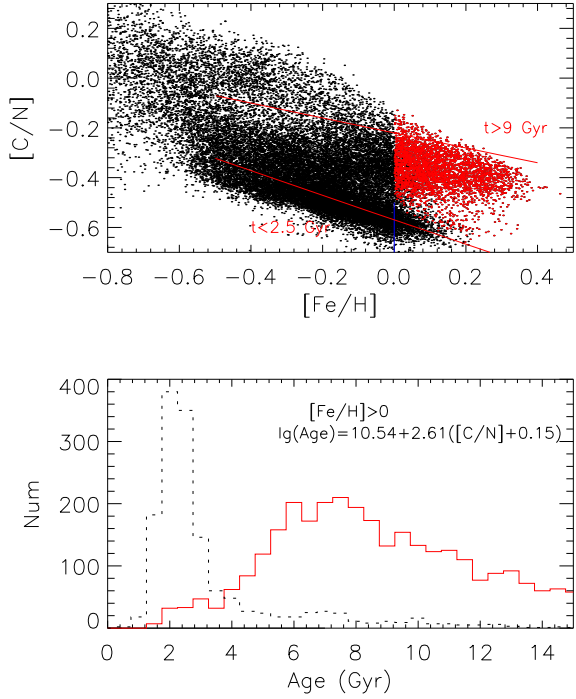}
\caption{Upper: The $\feh$ versus $\cn$ diagram with
the two red lines from \cite{Hasselquist19} corresponding the
ages of 2.5 Gyr and 9 Gyr based on open clusters.
Lower: The age histograms of low (black line) and high $\mgfe$ 
(red line) metal rich stars at $\feh>0$
based on the revised calibration of 
$log(Age[yrs])=10.54+2.61 (\cn+0.15)$ (from the original
one by \cite{Casali19}).}
\end{figure}

\subsection{The age-sensitive abundances}
Fig.~5 shows histograms of abundance ratios for six elements (Mg, C, O, Al, Si, Ti)
between the special and the comparison
groups. Besides $\mgfe$, other $\alpha$-related elements, $\ofe$ and $\tife$, as well as $\alfe$ in the former are overabundant.
Since these ratios are all age sensitive \citep{Delgado19}, it further
confirms the age separation between the two groups.
Meanwhile, the special group has wider abundance distributions
than the comparison group for all age-sensitive elements,
indicating a larger age range than the comparison group.
Note that $\sife$ distribution of the special group
overlaps significantly with the comparison group probably due to the
uncertainty of Si abundances in both the LAMOST-MRS-SPCANet DR7 and
APOGEE DR14 dataset. Actually, \cite{Zasowski19} revealed the presence
of abundance trends with temperature for Si and Ca elements.
We do not include $\cafe$ in the comparison not only because it 
has large uncertainties but also because it is not
age-sensitive \citep{Delgado19}.

The deviations in the peak of these histograms between the two groups 
are 0.10 dex for $\mgfe$, 0.16 dex for $\cfe$ and 0.12 dex for $\ofe$.
The overabundant $\mgfe$ of 0.10 dex is similar as that of 
the metal rich component in the bulge region with $R<3.5$ kpc 
by \cite{Rojas-Arriagada19}.
The overabundant $\ofe$ of $0.12$ dex with respect to the
comparison group is close to the average value of bar/bulge 
by \cite{Bovy19} (their Fig.~5.). 

\begin{figure}
\includegraphics[scale=1.0]{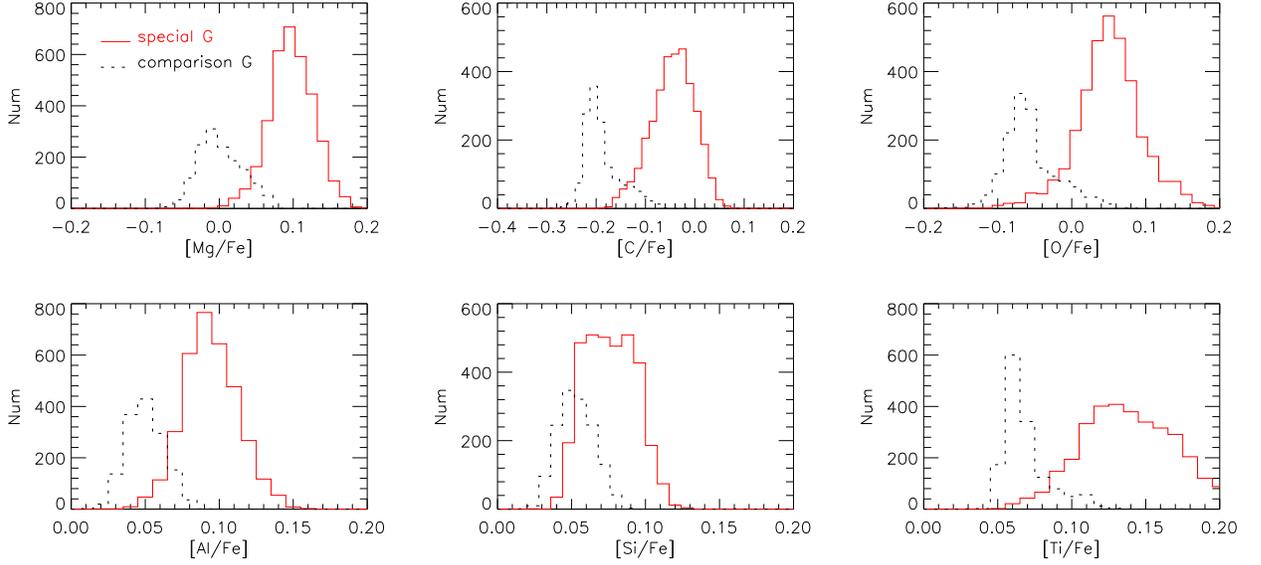}
\caption{The comparison of histograms of abundance ratios
for Mg,C,O,Al,Si,Ti elements between the special and the comparison groups.}
\end{figure}

\subsection{The $R$, $|Z|$ and $\feh$ histograms} 
The R, $|Z|$ and $\feh$ histograms between
the special and the comparison groups are presented
in Fig.~6. It shows that both groups locate around 9 kpc with
the former extending towards lower values until 7 kpc.
The comparison group has $|Z|<0.2$ kpc and $\feh<0.1$
in consistency with the origin of the young thin disk population,
while the special group has main contributions from 
stars with $|Z|>0.2$ kpc and with more extending metallicity 
towards $\feh\sim 0.3$ dex.

If the comparison group represents the local-born thin disk population
as indicated by young age and low $|Z|<0.2$ kpc, then the local ISM 
in the solar neighborhood is enriched to $\feh\sim 0.1$ at
3 Gyr ago. The local ISM metallicity should be significantly
lower at 7 Gyr ago when stars of the special group were born.
The fact that this intermediate-old special group has even higher
 metallicity than the young comparison group
highly suggests that it does not
belong to the local region but comes from a high-SFR region, e.g. the bulge.
According to \cite{Rojas-Arriagada19}, the bulge region
does show a high $\mgfe$ sequence extending to
super solar metallicity. Interestingly, they found that this metal rich 
component has 
an upper limit of $|Z|\sim 1.0$ kpc, similar as the $|Z|$ distribution 
of the special group (until 1.2 kpc) in our sample.
In addition, the X-shaped structure of the bulge
has a vertically extended feature to $|Z|\sim 1.2$ kpc \citep{Barbuy18}.
It seems that this special group with similar $|Z|$ extension
is related with the X-shaped structure of the bulge region.

\begin{figure}
\includegraphics[scale=1.0]{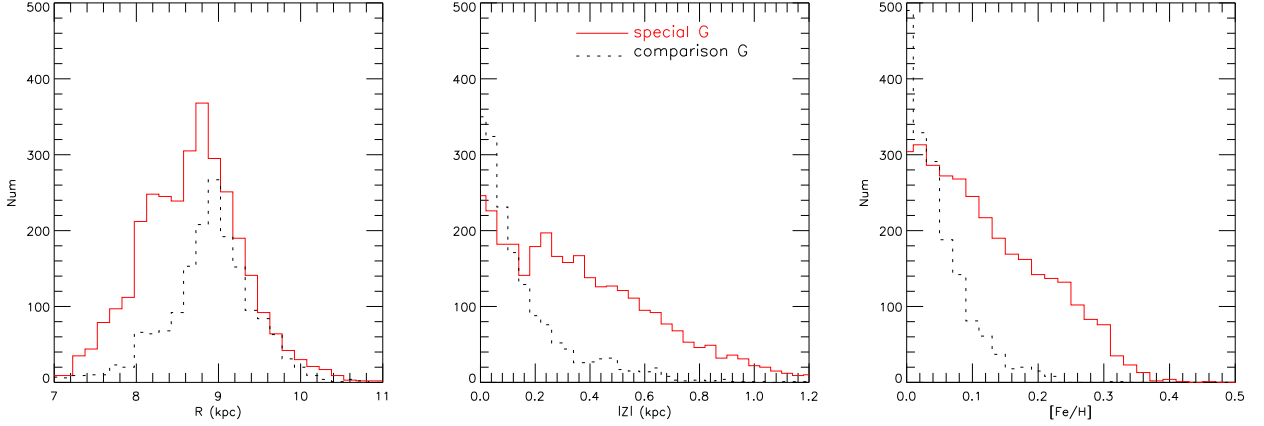}
\caption{The $R$, $|Z|$ and $\feh$ histograms of the special and the comparison groups.}
\end{figure}

\subsection{The kinematic and orbital parameters}
Fig.~7 presents the $R$ versus $\Vphi$ diagrams for the comparison
and the special groups as well as the thick disk represented by
stars with $|Z|>2$ kpc. Two dense clumps
are clear for the former and they become slightly sparse for
the latter, while there is almost no clump for stars with $|Z|>2$ kpc.
The velocity of the special group is more like
the thin disk as most stars distributing above the red dash line 
of $\Vphi\sim200 \kmprs$. Stars from the selected thick disk at 
$|Z|>2$ kpc show a wide distribution and
more stars have $\Vphi<200\kmprs$ than stars with $\Vphi>200\kmprs$.
The $V_R$ and $W_{LSR}$ velocities have the same features as $\Vphi$.
The low $\Vphi$ dispersion of this special group indicates its
kinematic colder than the group of high $|Z|$ thick disk. This is
consistent with the metal-rich component (mean $\feh=0.15$)
of the bulge in \cite{Ness13} who found this component
is kinematically cold and has the thin-disk characteristics, 
typical for the outer bar/bulge as shown in \cite{Bovy19}.

The distributions of eccentricity, $R-R_g$ and $R_{apo}-R_{peri}$ 
are similar for both groups as presented in Fig.~8. There is
a low eccentricity distribution of $e<0.3$, $|R-R_g|<2$ kpc but
quite a broad distribution of radial excursion distance 
of $R_{apo}-R_{peri}$ (from 1 to 5 kpc). Again, a
small fraction of stars of the special group extend to 
higher values in these histograms, which indicates a outward
displacement due to secular evolution.
In short, the special group is kinematically cold, similar as the
thin disk population, and has the same $|Z|$ distribution as 
the metal-rich component of the outer bulge and the X-shaped bar.

\begin{figure}
\includegraphics[scale=1.0]{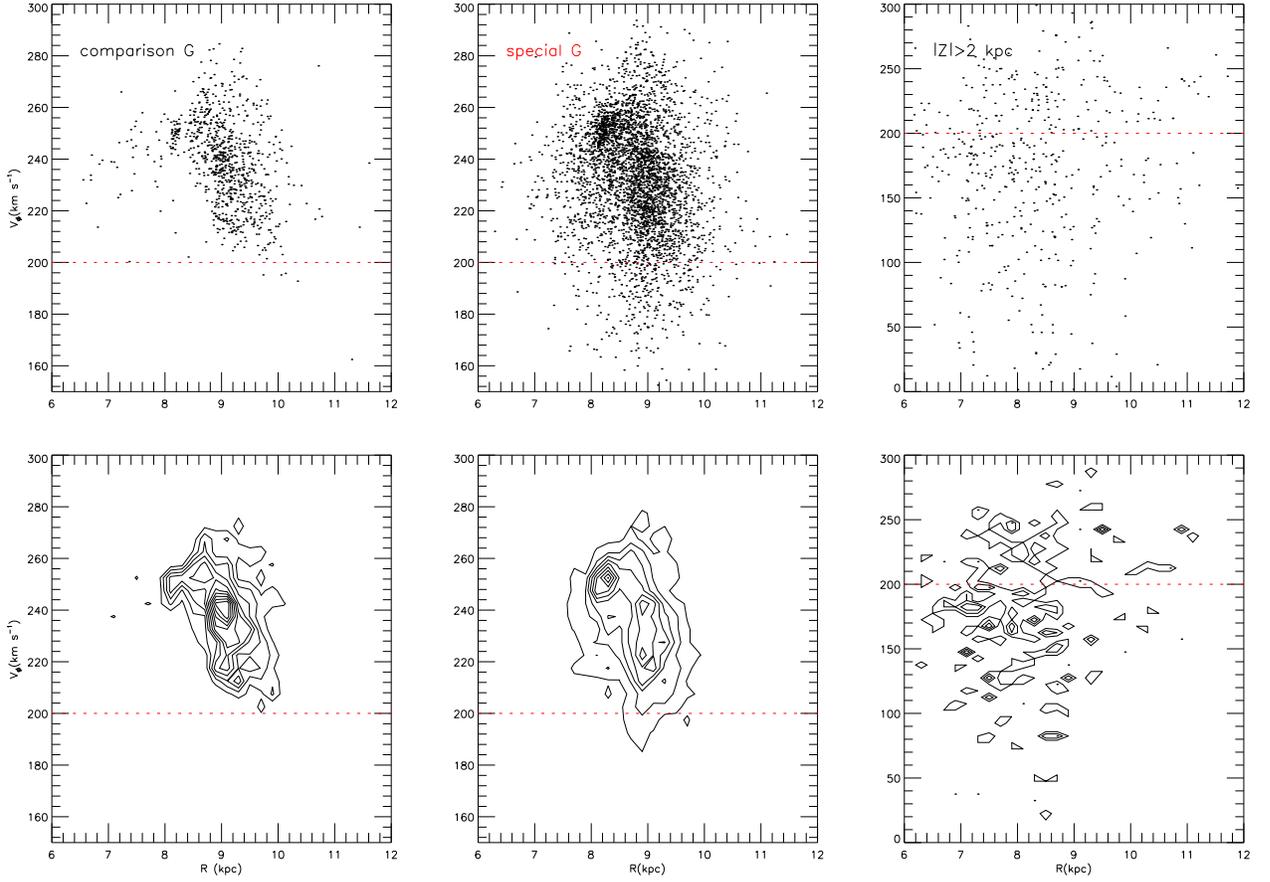}
\caption{The $R$ versus $\Vphi$ diagrams and the contour maps for
the comparison (left), the special (middle) groups and the thick disk
stars with $|Z|>2$ kpc (right).}
\end{figure}

\begin{figure}
\includegraphics[scale=1.0]{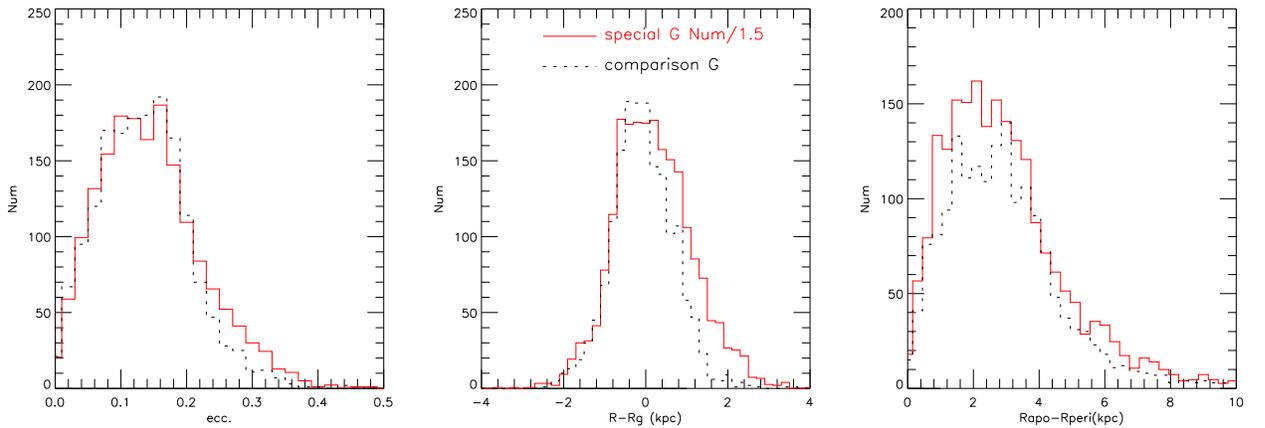}
\caption{The distributions of eccentricity, $R-R_g$ and $R_{apo}-R_{peri}$
for the special and the comparison groups.}
\end{figure}

\section{A tentative interpretation on the origin of the special group}
Based on the above properties (age, Galactic location, kinematics
and chemistry), we propose a tentative interpretation on the origin
of the special group by investigating how it connects 
with the bulge, where it comes from (the inner or outer region of the bulge), and how its stars
stray from their birth sites to present locations.

In the past few years, some observational and theoretical works have discussed the connection between these special stars and the inner disk. Using the Gaia ESO survey, \cite{Kordopatis15} pointed out that the metallicity of high-$\alpha$ stars in the solar neighborhood probably up to $0.2$ dex, and discussed that these stars may be formed in the inner most region of the Milky Way. However, due to the lack of reliable ages of these stars, they could not draw reliable conclusions. 
\cite{Minchev13} showed that most of the stars with super solar metallicity were born in the inner disk according to the chemodynamical simulation, and almost all the stars with metallicity of $0.25-0.6$ dex in the solar neighbourhood came from $3-5$ kpc. 
\cite{Minchev18} estimated the birth radius of field stars by a semi-empirical approach, and calculated that the birth radius of high-$\alpha$-metal-rich stars in their samples taken by the HARPS instrument is basically less than $5$ kpc (see their Fig.~9). 

According to \cite{Fragkoudi18},
the bulge is slightly metal-poor with $\feh=-0.15$ dex (see their Fig.~13)
in the innermost regions
($R<1$ kpc), while the most metal rich component has the mean metallicity
of $\feh=+0.1$ dex in the outer bulge where the bar is strong.
However, \cite{Wegg19} suggested that stars in the central bar (also
the central bulge) are more metal-rich than their surrounding region and 
thus suggested an opposite metallicity structure of the bulge.
Based on APOGEE and Gaia surveys, \cite{Bovy19}
presented unprecedented detailed maps of the kinematics, elemental
abundances, and age of the bulge region, and found that
the central bar tend to be slightly metal poor than the bulge
region outside the bar, which is consistent with the result of \cite{Fragkoudi18}. 
In particular, they suggested that the highest metallicity region is
in the outermost region of the bar/bulge, i.e.
$3-5$ kpc, where the age has a peak of $\sim$ 7 Gyr and
an enhancement $\ofe \sim0.08$ dex (with
respect to solar radius) as shown in their figure 5.
All these properties are found for the special group in our sample,
and thus we suggest that it may originate
from $R=3-5$ kpc, rather than the central bar/bulge ($R<3$ kpc).

The bar/bulge shows an X-shape (or boxy/peanut shape) structure
in the X-Z diagram \citep{Barbuy18}, which has an extension to $\sim 1.2$ kpc 
in the $|Z|$ direction.
According to \cite{DiMatteo16}, the X-shape structure
is found to be more prominent
in the metal-rich population than other components. In other words,
stars in the most metal rich component of the bulge have
a large extension in vertical distance, reaching 1.2 kpc at the outer 
region of $3-5$ kpc. The similar $|Z|$ extension to 1.2 kpc for 
the special group in our work supports the 
suggestion that its stars come from the outer bar/bulge region 
at $R=3-5$ kpc and $0<|Z|<1.2$ kpc.

Finally, it is interesting to investigate how stars in this special group
move from their birth sites of $3-5$ kpc to present locations peaking
at $8-9$ kpc (see Fig.~6). The low eccentricity (peaking at
$0.1-0.2$), small $|R-R_g|$ ($<2$ kpc) and kinematical cold (small
velocity dispersion) all indicate that it is unlikely 
that blurring alone could cause a distance displacement of $3-5$ kpc in 
the radial direction without inducing significant velocity dispersions.
Churning (i.e. radial migration) is, however, a promising mechanism
to bring these stars to present locations in the sense that
their birth sites of $3-5$ kpc are close to the corotation radius (CR)
of the Galactic bar, which is located at 4.7 kpc \citep{Minchev13} or 
5.5 kpc \citep{Bovy19},
where radial migration is most intense.
According to \cite{Minchev13}, the incorporating spiral arms and bar structure
could provide a very powerful stellar radial migration,
leading to the migration distance as large as 5 kpc.
Moreover, radial excursions between peri-center and apo-center
distances ($R_{apo}-R_{peri}$) are of $1-5$ kpc (Fig.~8), which allows for a
radial displacement of these stars outwards to larger distances.
The low eccentricity of the special group
is consistent with the scenario that these stars stray
from the outer bar/bulge at $R\sim3-5$ kpc and
moved outwards by radial migration, which alters the guiding radius without
changing the eccentricity. 
If this is the case, its thin-disk characteristics, i.e.
kinematic-cold, intermediate-age and mildly-enhanced $\mgfe$ ratio,
would provide a support for the bar/bulge formation
from the thin disk \citep{DiMatteo16}.

\section{Result}
Based on LAMOST-MRS-SPCANet DR7 data, we selected a sample
of stars with $S/N>20$ and stellar parameters in the ranges 
of $4200<\teff < 5200\,K$, $2.6<logg < 3.3$ and $-0.8<\feh<0.6$ dex.
In the $\feh$ versus $\mgfe$ diagram, 
two distinct groups are detected at super solar metallicity,
a special group with enhanced $\mgfe$ of 0.1 dex and a comparison group
with solar $\mgfe$ ratios.

Stellar ages are estimated from $\cn$ ratios, and this special group
corresponds to an intermediate-age population of $5-9$ Gyr, while the 
comparison group is a young population peaking at $2-3$ Gyr.
The special group shows wide distributions with
significant extension towards
small $R$, large $|Z|$ (until 1.2 kpc) and high $\feh$ as compared with 
the comparison group. The age-sensitive elements, $\ofe$, $\cfe$, $\alfe$ and
$\tife$ in the special group are also overabundant as well as 
the enhanced $\mgfe$ and $\cn$ ratios.
Both groups are thin-disk like in the $R$ versus $\Vphi$ diagrams
and are kinematic cold as indicated by low velocity dispersions.
We connect this special group with the outer bar/bulge region
based on similar properties, i.e. the extending distributions 
of high $|Z|$ and high $\feh$, slightly overabundant $\afe$ ratios,
thin-disk like kinematics and low velocity dispersions.

By combining the above properties, we propose
a tentative interpretation on the origin of the special group.
It was born at the outer bar/bulge of
$R\sim 3-5$ kpc, where it is the most metal-rich component 
with an extending $|Z|$ distribution to 1.2 kpc,
typical for the X-shape structure at the outer bar/bulge. 
Radial migration induced
by the coupling between the bar and the spiral arms \citep{Minchev13}
could bring its stars from the birth sites of $3-5$ kpc
to present locations. 
If the bulge origin of this special group
is true, its kinematic-cold, intermediate-age and mildly-enhanced $\mgfe$
would provide a support for the scenario that the bar/bulge is formed
from the thin disk population as suggested by \cite{DiMatteo16}.
The future 2-m Chinese Space Survey Telescope (CSST) project \citep{Zhan11} will provide a good opportunity to investigate the properties of the bulge
and the mechanism of radial migration by finding a large amount of
super solar metallicity stars in the inner disk. The preparation of
the usage of data from the CSST project for this investigation is desirable,
as already started for the study of the merging history by \cite{Zhao21}
via the chemical abundances of halo stars.

\begin{acknowledgements}
This study is supported by the National Natural Science Foundation of China under grants No. 11988101, 11625313, 11890694, 11973048, 11927804, the 2-m Chinese Space Survey Telescope project and the National Key R\&D Program of China No. 2019YFA0405502. This work is also supported by the Astronomical Big Data Joint Research Center, co-founded by the National Astronomical Observatories, Chinese Academy of Sciences and the Alibaba Cloud. 

The Guo Shou Jing Telescope (the Large Sky Area Multi-Object Fiber Spectroscopic Telescope LAMOST) is a National Major Scientific Project built by the Chinese Academy of Sciences. Funding for the project has been provided by the National Development and Reform Commission. The LAMOST is operated and managed by the National Astronomical Observatories, Chinese Academy of Sciences.

This work has made use of data from the European Space
Agency (ESA) mission Gaia (https://www.cosmos.esa.
int/gaia), processed by the Gaia Data Processing and Analysis
Consortium (DPAC, https://www.cosmos.esa.int/
web/gaia/dpac/consortium). Funding for the DPAC has
been provided by national institutions, in particular the institutions
participating in the Gaia Multilateral Agreement.
\end{acknowledgements}

\end{document}